\documentclass[12pt]{article}

\usepackage{axodraw}
\usepackage{epsfig}
\usepackage{rotating}
\def\ie{{\it i.e.}}
\def\lsim{\:\raisebox{-0.5ex}{$\stackrel{\textstyle<}{\sim}$}\:}
\def\gsim{\:\raisebox{-0.5ex}{$\stackrel{\textstyle>}{\sim}$}\:}
\def\bmaT{\left(\begin{array}{ccc}}
\def\emaT{\end{array}\right)}
\def\bma{\left( \begin{array} }
\def\ema{\end{array} \right)}

\begin{document}
{\flushright
\hspace{30mm} \\
\hspace{30mm} November 2007 \\
}
\vspace{1cm}

\begin{center}
{\Large\sc {\bf Effect of Charged Scalar Loops on Photonic Decays\\
of a Fermiophobic Higgs}}
\vspace*{3mm}
\vspace{1cm}

{\large {A.G. Akeroyd}$^{a,b}$, {Marco A. D\'{\i}az}$^c$, 
{Maximiliano A. Rivera}$^c$}\\
\vspace{1cm}
{\sl
a: Department of Physics, National Cheng Kung University, 
Tainan 701, Taiwan\\
\vspace*{0.2cm}
b: National Center for Theoretical Sciences, Taiwan\\
\vspace*{0.2cm}
c: Departamento de F\'{\i}sica, Universidad Cat\'olica de Chile,\\
Avenida Vicu\~na Mackenna 4860, Santiago, Chile 
\vspace*{0.2cm}
}
\end{center}

\vspace{2cm}

\begin{abstract}
\noindent

Higgs bosons with very suppressed couplings to fermions
(``Fermiophobic Higgs bosons'', $h_f$) can decay to two photons
($\gamma\gamma$) with a branching ratio significantly
larger than that expected for the Standard Model Higgs boson
for $m_{h_f}<150$ GeV.
Such a particle would give a clear signal at the LHC and can arise in the 
Two Higgs Doublet Model (type~I) in which $h_f\to \gamma\gamma$
is mediated by $W^\pm$ and charged Higgs boson ($H^\pm$) loops.
We show that the $H^\pm$ loops can cause both constructive and 
destructive contributions with a magnitude considerably larger than the 
anticipated precision in the measurement of the photonic
decay channel at future hadron and lepton colliders.


\end{abstract}

\newpage

\section{Introduction}

Neutral Higgs bosons with very suppressed couplings to fermions --
``fermiophobic Higgs bosons'' ($h_f$) \cite{Weiler:1987an}-- may arise in
specific versions of the Two Higgs Doublet Model (2HDM)
\cite{Haber:1979jt, Gunion:1989we} or in models with Higgs triplets
\cite{Georgi:1985nv}. Such a $h_f$ would decay dominantly to two photons,
$h_f\to \gamma\gamma$, for $m_{h_f}\lsim 95$ GeV or to 
two massive gauge bosons,
$h_f\to VV^{(*)}$, ($V=W^\pm,Z$) for $m_{h_f}\gsim 95$ GeV
\cite{Stange:1994ya,Diaz:1994pk}.  The large branching ratio for 
$h_f\to \gamma\gamma$ would provide a very clear experimental signature, 
and observation of such a particle would strongly constrain the 
possible choices of the underlying Higgs sector
\cite{Stange:1994ya}-\cite{Brucher:1999tx}.

Experimental searches for $h_f$ have been performed at 
the CERN Large Electron Positron Collider (LEP) and 
the Fermilab Tevatron. Lower limits on $m_{h_f}$ have 
been derived in the context of a benchmark model which assumes
that the coupling $h_fVV$ is equal to the Standard Model (SM)
Higgs boson coupling $\phi^0VV$, and that all fermion branching 
ratios are exactly zero. 
We will refer to this model as {\sl Benchmark F}. Lower
bounds of the order $m_{h_f}\gsim 100$ GeV have been obtained by the LEP
collaborations OPAL\cite{Abbiendi:2002yc}, DELPHI\cite{Abreu:2001ib},
ALEPH\cite{Heister:2002ub}, and L3\cite{Achard:2002jh}, utilizing the channel
$e^+e^-\to h_fZ$, $h_f\to \gamma\gamma$. A search in the
complementary channel $e^+e^-\to A^0h_f$ was performed by 
two LEP collaborations, OPAL \cite{Abbiendi:2002yc} and 
DELPHI \cite{Abreu:2001ib} and
ruled out the region $m_A+m_{h_f}< 160$ GeV. 

At the Tevatron Run I, the limits on
$m_{h_f}$ from the D\O\ and CDF collaborations are respectively 78.5 GeV
\cite{Abbott:1998vv} and 82 GeV \cite{Affolder:2001hx} at $95\%$ C.L., using
the mechanism $qq'\to V^*\to h_fV$, $h_f\to \gamma\gamma$, with the dominant
contribution coming from $V=W^\pm$. For an integrated luminosity of 2
fb$^{-1}$ (which has been attained as of June 2007) Run II can extend the 
coverage of $m_{h_f}$ in {\sl Benchmark F} model
slightly beyond that of LEP \cite{Mrenna:2000qh}-\cite{Landsberg:2000ht}. In
addition, Run II is sensitive to the region $110 \,{\rm GeV} < m_{h_f} < 160
\,{\rm GeV}$ and $B(h_f\to \gamma\gamma)>4\%$ which could not be probed at
LEP.  A preliminary search in the inclusive $2\gamma$ channel has been
performed with $0.19 \,$fb$^{-1}$ of Run II data 
\cite{Melnitchouk:2005xv, D0result}.
A complementary production mechanism which 
is exclusive to a hadron collider is the process 
$qq'\to H^\pm h_f$ \cite{Akeroyd:2003bt,Akeroyd:2003xi}.
Since the branching ratio for $H^\pm\to h_f W^{(*)}$
can be very large in models with fermiophobia 
\cite{Akeroyd:1998dt} this mechanism can give rise to double $h_f$
production and hence a multiphoton signature.  
The sensitivity of the Tevatron Run II to this channel 
was studied in \cite{Akeroyd:2005pr} and recently a 
search was performed by the D\O\ collaboration in the 
$3\gamma$ channel \cite{Landsberg:2007mc}.

In {\sl Benchmark F} the decay $h_f\to \gamma\gamma$ is assumed to be
mediated solely by $W^\pm$ loops. In this paper we study the 
effect of charged scalar loops ($H^\pm$) on $B(h_f\to \gamma\gamma)$ in the
context of the 2HDM (type I) and discuss the impact on the
current and future searches for $h_f$ at the Tevatron and LHC.
Our work is organized as follows: in section 2 the fermiophobic limit
of the 2HDM (type I) is introduced; in section 3 the effect of
the charged scalar loops on $h_f\to \gamma\gamma$ is discussed;
the numerical results for $B(h_f\to \gamma\gamma)$
are contained in section 4 with conclusions given
in section 5.

\section{Fermiophobic Higgs bosons}

In this section we briefly review the properties of $h_f$
in the 2HDM (type I).  For a detailed
introduction the reader is referred to \cite{Diaz:1994pk}, 
\cite{Akeroyd:1996hg}-\cite{Brucher:1999tx}.

\subsection{2HDM (Type I)}    

If $\Phi_1$ and $\Phi_2$ are two Higgs $SU(2)$ doublets with 
hypercharge $Y=1$, the most general $SU(2)_L\times U(1)_Y$ gauge invariant
scalar potential is \cite{Gunion:2002zf}:
\begin{eqnarray}
V&=&m_{11}^2\Phi_1^{\dagger}\Phi_1+m_{22}^2\Phi_2^{\dagger}\Phi_2-
\left( m_{12}^2\Phi_1^{\dagger}\Phi_2+{h.c.}\right)+
{\textstyle{1\over2}}\lambda_1\left(\Phi_1^{\dagger}\Phi_1\right)^2
\nonumber\\ &&
+{\textstyle{1\over2}}\lambda_2\left(\Phi_2^{\dagger}\Phi_2\right)^2
+\lambda_3\left(\Phi_1^{\dagger}\Phi_1\right)
\left(\Phi_2^{\dagger}\Phi_2\right)
+\lambda_4\left(\Phi_1^{\dagger}\Phi_2\right)
\left(\Phi_2^{\dagger}\Phi_1\right)
\label{scalar_pot}\\ &&
+\left\{{\textstyle{1\over2}}\lambda_5
\left(\Phi_1^{\dagger}\Phi_2\right)^2
+\left[\lambda_6\left(\Phi_1^{\dagger}\Phi_1\right)+
\lambda_7\left(\Phi_2^{\dagger}\Phi_2\right)\right]\Phi_1^{\dagger}\Phi_2
+{h.c.}\right\} \; .
\nonumber
\end{eqnarray}
If the discrete symmetry $\Phi_1\rightarrow-\Phi_1$ is imposed 
one has $\lambda_6=\lambda_7=0$. However, the term proportional to
$m_{12}^2$ can remain as a soft violation of the above discrete symmetry
and still ensure that Higgs-mediated tree-level flavour changing neutral 
currents are absent \cite{Gunion:1989we}.
Note that the above 2HDM potential contains one more free parameter than
those studied in Refs.~\cite{Barroso:1999bf,Brucher:1999tx}.
We assume that all the scalar potential parameters are real.

The scalar potential in eq.~(\ref{scalar_pot}) 
breaks $SU(2)_L\times U(1)_Y$ down 
to $U(1)_{em}$ when the two Higgs doublets acquire vacuum expectation 
values
\begin{equation}
\langle \Phi_1 \rangle = \frac{1}{\sqrt{2}}
\left(\matrix{0 \cr v_1}\right)
\,,\qquad
\langle \Phi_2 \rangle = \frac{1}{\sqrt{2}}
\left(\matrix{0 \cr v_2}\right)
\end{equation}
which must satisfy the experimental constraint
$m_Z^2=\frac{1}{2}(g^2+g'^2)v^2$, with 
$v^2=(v_1^2+v_2^2)\approx(246~ \mathrm{GeV})^2$.
The minimization conditions that define the vacuum expectation values
in terms of the parameters of the potential (setting $\lambda_6=\lambda_7=0$)
are
\begin{eqnarray}
t_1&=&m_{11}^2 v_1-m_{12}^2 v_2+\textstyle{1\over2}\lambda_1v_1^3
+\textstyle{1\over2}(\lambda_3+\lambda_4+\lambda_5)v_1v_2^2\,=\,0
\nonumber\\
t_2&=&m_{22}^2 v_2-m_{12}^2 v_1+\textstyle{1\over2}\lambda_2v_2^3
+\textstyle{1\over2}(\lambda_3+\lambda_4+\lambda_5)v_1^2v_2\,=\,0
\label{tadpoles}
\end{eqnarray}
from which $m_{11}^2$ and $m_{22}^2$ can be solved in favour of $m_Z^2$
and $\tan\beta\equiv v_2/v_1$.

After using the minimization conditions, the neutral CP-odd Higgs mass 
matrix can be written as
\begin{equation}
{\bf M}^2_A=\left(\matrix{
m_{12}^2t_{\beta}-\lambda_5v^2s_{\beta}^2 &
-m_{12}^2+\lambda_5v^2s_{\beta}c_{\beta} \cr
-m_{12}^2+\lambda_5v^2s_{\beta}c_{\beta} &
m_{12}^2/t_{\beta}-\lambda_5v^2c_{\beta}^2
}\right)
\end{equation}
and is diagonalized by a rotation in an angle $\beta$. For simplicity we use
the following notation,
$s_\beta = \sin\beta$, $c_\beta = \cos\beta$, and $t_\beta =\tan
\beta$. The mass matrix 
${\bf M}^2_A$ has a zero eigenvalue corresponding to the
neutral Goldstone boson while its second eigenvalue is the mass of the
physical CP-odd Higgs boson $A^0$,
\begin{equation}
m_A^2=\frac{m_{12}^2}{s_{\beta}c_{\beta}}-\lambda_5v^2
\label{ma}
\end{equation}
The charged Higgs mass matrix is given by
\begin{equation}
{\bf M}^2_{H^{\pm}}=\left(\matrix{
m_{12}^2t_{\beta}-\frac{1}{2}(\lambda_4+\lambda_5)v^2s_{\beta}^2 &
-m_{12}^2+\frac{1}{2}(\lambda_4+\lambda_5)v^2s_{\beta}c_{\beta} \cr
-m_{12}^2+\frac{1}{2}(\lambda_4+\lambda_5)v^2s_{\beta}c_{\beta} &
m_{12}^2/t_{\beta}-\frac{1}{2}(\lambda_4+\lambda_5)v^2c_{\beta}^2
}\right)
\end{equation}
which also is diagonalized by a rotation in an angle $\beta$. It has a 
zero eigenvalue corresponding to the charged Goldstone boson, and the 
charged Higgs mass is
\begin{equation}
m_{H^{\pm}}^2=m_A^2+\frac{1}{2}(\lambda_5-\lambda_4)v^2  \; .
\end{equation}
Here we see that the charged and the CP-odd Higgs masses are independent
parameters, as opposed to supersymmetry, where the mass squared difference 
is equal to $m_W^2$ at tree-level.

The neutral CP-even Higgs mass matrix is given by
\begin{equation}
{\bf M}^2_{H^0}=\left(\matrix{
m_A^2s_{\beta}^2+\lambda_1v^2c_{\beta}^2+\lambda_5v^2s_{\beta}^2  &
-m_A^2s_{\beta}c_{\beta}+(\lambda_3+\lambda_4)v^2s_{\beta}c_{\beta} \cr
-m_A^2s_{\beta}c_{\beta}+(\lambda_3+\lambda_4)v^2s_{\beta}c_{\beta} &
m_A^2c_{\beta}^2+\lambda_2v^2s_{\beta}^2+\lambda_5v^2c_{\beta}^2
}\right)
\end{equation}
and the two eigenvalues are the masses of the neutral CP-even Higgs
bosons $h^0$ and $H^0$. It is diagonalized by an angle $\alpha$ defined by
\begin{equation}
\sin2\alpha={{\left[-m_A^2+(\lambda_3+\lambda_4)v^2\right]s_{2\beta}}
\over
{\sqrt{
\left[(m_A^2+\lambda_5v^2)c_{2\beta}-\lambda_1v^2c_{\beta}^2+\lambda_2v^2s_{\beta}^2
\right]^2+\left[m_A^2-(\lambda_3+\lambda_4)v^2\right]^2s^2_{2\beta}
}}} \; .
\label{alpha}
\end{equation}
A necessary condition for fermiophobia is the imposition of 
the mentioned discrete symmetry 
$\Phi_1\to -\Phi_1$ which forbids $\Phi_1$ coupling to the fermions.  
This model is usually called ``Type I'' \cite{Haber:1979jt}.  However, 
fermiophobia is only partial due to the mixing in the CP--even neutral Higgs 
mass matrix, which is diagonalized by the mixing angle $\alpha$, and
both CP--even mass eigenstates $h^0$ and $H^0$ acquire a coupling to the 
fermions.  
The fermionic couplings of the lightest CP--even Higgs $h^0$ take the form
$h^0f\overline f \sim \cos\alpha/\sin\beta$, where $f$ is any fermion.
Small values of $\cos\alpha$ would strongly suppress the fermionic 
couplings, and in the limit $\cos\alpha \to 0$ the coupling 
$h^0f\overline f$ would vanish, giving rise to 
complete fermiophobia at tree-level.\footnote{The limit 
$\sin\alpha \to 0$ is studied in Ref.\cite{Phalen:2006ga}
and can give rise to a very suppressed $B(h^0\to \gamma\gamma$)
in the 2HDM (Model~I).}
 From eq.~(\ref{alpha}) this is 
achieved if
\begin{equation}
m_A^2=(\lambda_3+\lambda_4)v^2 \; .
\label{ferm_condition}
\end{equation}
Despite this extra constraint, the parameters $m_A$, $m_{H^{\pm}}$, and
$\tan\beta$ are still independent parameters in this model.
However, at the one-loop level, $h_f$ can couple to fermions
via loops involving vector bosons and other Higgs 
bosons (for a discussion see 
\cite{Diaz:1994pk,Barroso:1999bf,Brucher:1999tx,Akeroyd:2003xi}). 
Thus in general one would expect a small
$B(h_f\to f\overline f)$ even if fermiophobia were exact at tree-level.
Despite this, it is conventional and convenient to define an extreme $h_f$ 
in which all branching ratios to fermions are set to zero. 
This can be arranged by defining a vanishing renormalized $\alpha$. 

\subsection{Fermiophobic Higgs boson branching ratios}

Expressions for the branching ratio $B(h_f\rightarrow\gamma\gamma)$
can be found in ref.~\cite{Gunion:1989we}. In {\sl Benchmark F} this decay
is assumed to be mediated solely by $W$ boson loops,
\begin{center}
\vspace{-85pt} \hfill \\
\begin{picture}(300,140)(0,23) 
\DashLine(20,30)(60,30){3}
\Photon(60,30)(90,60){3}{6.5}
\Photon(90,0)(60,30){3}{6.5}
\Photon(90,60)(90,0){3}{8}
\Photon(90,60)(130,60){3}{6.5}
\Photon(90,0)(130,0){3}{6.5}
\Text(20,40)[]{$h_f$}
\Text(140,60)[]{$\gamma$}
\Text(140,0)[]{$\gamma$}
\Text(107,30)[]{$W^{\pm}$}
\DashLine(170,30)(210,30){3}
\PhotonArc(230,30)(20,0,360){3}{18}
\Photon(250,30)(280,60){3}{6.5}
\Photon(250,30)(280,0){3}{6.5}
\Text(170,40)[]{$h_f$}
\Text(290,60)[]{$\gamma$}
\Text(290,0)[]{$\gamma$}
\Text(230,60)[]{$W^{\pm}$}
\end{picture}  
\vspace{30pt} \hfill \\
\end{center}
with a SM-strength Higgs-$W$-$W$ coupling, although in the 2HDM a factor 
of $\sin(\beta-\alpha)$ must be included.  
In the fermiophobic limit ($\cos\alpha=0$) the $h_fWW$ coupling
($g_{h_fWW}$) normalized to the SM $\phi_0WW$ coupling satisfies
$\sin(\beta-\alpha)\rightarrow-\cos\beta$. We call this scenario 
{\sl Benchmark B}. The decay 
rate for $h_f$ into two photons is as follows:
\begin{equation}
\Gamma(h_f \to \gamma\gamma)=
{\alpha^2g^2\over 1024\pi^3}{m^3_{h_f}\over m_W^2} 
|F_1\cos\beta|^2 
\label{hggWW}
\end{equation}
Here $g$ is the $SU(2)_L$ coupling constant, $\alpha$ is the
fine-structure constant, and $F_1=F_1(\tau)$ where
$\tau=4m^2_W/m^2_{h_f}$ is a phase space function given in 
\cite{Gunion:1989we}. 

\begin{figure}
\centerline{\protect\hbox{\epsfig{file=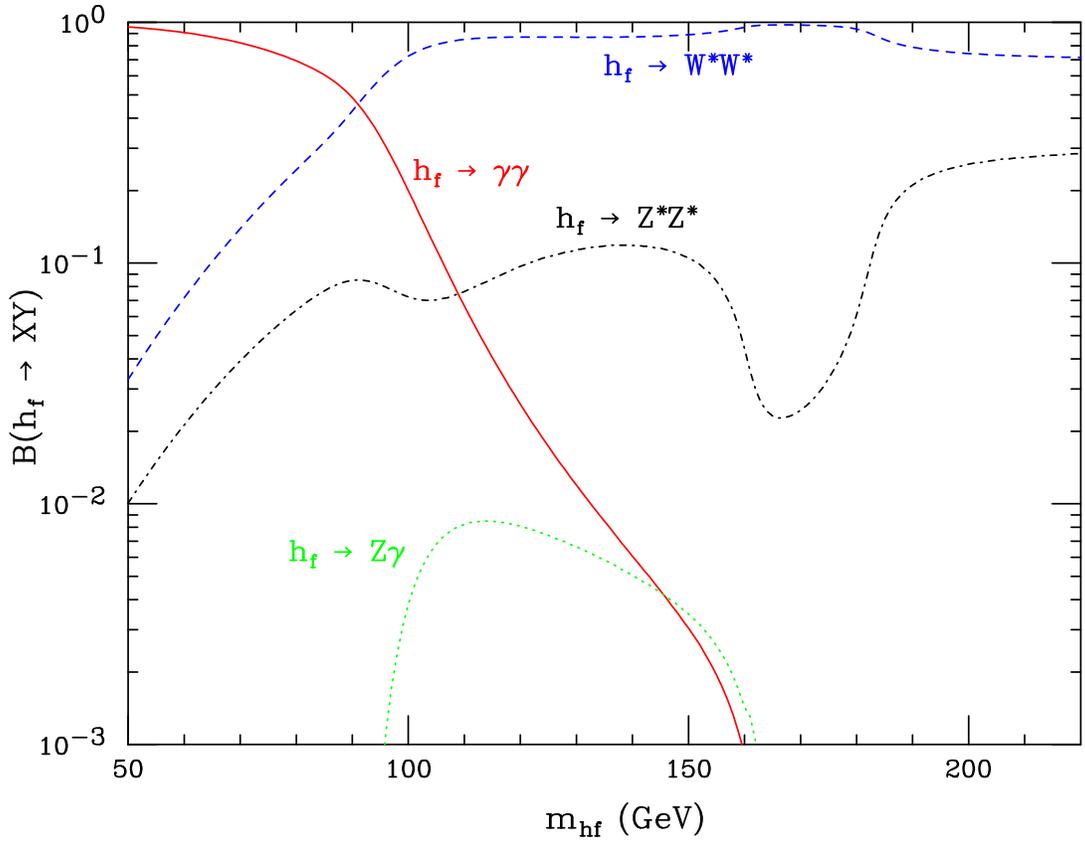,width=0.7\textwidth,angle=90}}}
\caption{\it $B(h_f\rightarrow \gamma \gamma,W^*W^*,Z^*Z^*, Z\gamma)$ as a 
function
of $m_{h_f}$ assuming all fermionic decays are absent (exact fermiophobic) in
the Benchmark Models F and B.}
\label{bench}
\end{figure}
For the sake of illustration, we depict in Fig.\ \ref{bench} the branching
ratios of a fermiophobic Higgs boson $h_f$ into $VV$ where $V$ can be 
either a $W$, $Z$ or $\gamma$
(see \cite{Stange:1994ya,Diaz:1994pk} for earlier versions
of the same figure).
In  Fig.\ \ref{bench} it is assumed that all $h_f$ 
couplings to fermions are absent and that $h_f\to \gamma\gamma$ 
is mediated solely by a $W$ boson loop.
One can see from
the figure that the loop induced decay mode $h_f \to \gamma\gamma$ is
dominant for $m_{h_f} \lsim 95$ GeV and drops below $0.1 \%$ for $h_f$
masses above 150 GeV. On the other hand, the decay channel $h_f \to
W^* W^*$ dominates for $m_{h_f} \gsim 95$ GeV, being close to 100\%
until the threshold for $h_f$ decay into two real $Z$'s is reached.
Note that the branching ratios in Fig.~\ref{bench} coincide for both
{\sl Benchmark B} and {\sl F}, although the decay rates and production
cross--sections are different.

\section{Charged scalar loop contributions to $h_f\to \gamma\gamma$}

In the 2HDM (type I) there are additional contributions to $h_f\to\gamma\gamma$
from singly charged scalar ($H^\pm$) loops 
\cite{Gunion:1989we,Barroso:1999bf,Brucher:1999tx}\footnote{See 
\cite{Ginzburg:2001wj} 
for studies of the effect of
charged scalar loops on the photonic decays of neutral Higgs bosons in the 
context of
the 2HDM (type II).} whose graphs are shown below:
\begin{center}
\vspace{-10pt} \hfill \\
\begin{picture}(300,140)(0,23) 
\DashLine(20,120)(60,120){3}
\DashLine(60,120)(90,150){3}
\DashLine(90,90)(60,120){3}
\DashLine(90,150)(90,90){3}
\Photon(90,150)(130,150){3}{6.5}
\Photon(90,90)(130,90){3}{6.5}
\Text(20,130)[]{$h_f$}
\Text(140,150)[]{$\gamma$}
\Text(140,90)[]{$\gamma$}
\Text(105,122)[]{$H^{\pm}$}
\DashLine(170,120)(210,120){3}
\DashCArc(230,120)(20,0,360){3}
\Photon(250,120)(280,150){3}{6.5}
\Photon(250,120)(280,90){3}{6.5}
\Text(170,130)[]{$h_f$}
\Text(290,150)[]{$\gamma$}
\Text(290,90)[]{$\gamma$}
\Text(233,150)[]{$H^{\pm}$}
\end{picture}  
\vspace{-40pt} \hfill \\
\end{center}
Such contributions, which are neglected in both 
{\sl Benchmark B} and {\sl F}, 
introduce a dependence on the details of the Higgs spectrum. 
The expression for $\Gamma(h_f\to \gamma\gamma)$ in 
eq.(\ref{hggWW}) is modified to the following:
\begin{equation}
\Gamma(h_f\to \gamma\gamma)={\alpha^2g^2\over 1024\pi^3}
{m^3_{h_f}\over m_W^2} 
|F_0 \, \tilde g_{h_fH^+H^-}-F_1\cos\beta |^2 
\label{withscalarloops}
\end{equation}
Here the dimensionless coupling $\tilde g_{h_fH^+H^-}$ is proportional to 
the trilinear 
coupling for the $h_fH^+H^-$ vertex, which is defined below.
$F_0$ is a phase space function  of ($m^2_{H^\pm}/m^2_{h_f}$),
\ie, the scalar analogue of $F_1$.
In the mass region of interest to us, $100 {\rm GeV} < m_{h_f} < 250$ GeV,
$F_0$ is considerably smaller than $F_1$ e.g., for $m_{h_f}< 2m_{H^\pm}$ 
one has $-1/3 > F_0 > -1$ while $7 < F_1 < 12$ for $m_{h_f}< 2m_W$.
However, the impact of the scalar loops can be significant
if there is some suppression for the $h_fWW$ coupling 
or if $\tilde g_{h_fH^+H^-}$ is sizeable. 
Hereafter we will refer to the scenario
where the charged Higgs boson loops are included
{\it and} the 2HDM $g_{h_fWW}$ coupling is used as {\sl Benchmark A}. 

The trilinear coupling $\tilde g_{h_fH^+H^-}$ in eq.~(\ref{withscalarloops}) 
is related to the coupling in the Lagrangian,
$L= g_{h_fH^+H^-}h_fH^+H^-+\dots$, by
\begin{equation}
\tilde g_{h_f H^+H^-}= -\frac{m_W}{gm_{H^+}^2}g_{h_f H^+H^-}
\label{trilinear_coup1}
\end{equation}
where,
\begin{equation}
g_{h_f H^+H^-}= c_\beta v\left[2\frac{m_{H^\pm}^2}{v^2}+s_\beta^2\lambda_1-
(1+s_\beta^2)\frac{m_A^2}{v^2}-(1+s_\beta^2)\lambda_5\right]
\label{trilinear_coup2}
\end{equation}
Eq.~(\ref{trilinear_coup2}) is obtained by imposing the
fermiophobic condition [eq.~(\ref{ferm_condition})]
on the expression for $g_{h_f H^+H^-}$ in the general
2HDM \cite{Gunion:2002zf}.
%
Note that at the Lagrangian level the trilinear coupling $g_{h_fH^+H^-}$
has dimensions of mass. 

Clearly the contribution of the $H^\pm$ loops depends  
on the details of the scalar potential.
The phase space function $F_0$ involves the scalar masses
$m_{h_f}$ and $m_{H^\pm}$, 
while $g_{h_fH^+H^-}$ is a function of
several Higgs potential parameters. 
Since the charged scalar contribution
may interfere destructively or constructively with
that of the $W$ loop (depending on the sign of
$g_{h_fH^+H^-}$), its main phenomenological effect on
the decay $h_f\to \gamma\gamma$ is to increase or decrease
B$(h_f\to \gamma\gamma)$ for a given $m_{h_f}$ with respect to that given 
in Fig.\ref{bench}. A first study of the effect of the
$H^\pm$ loops on B$(h_f\to \gamma\gamma)$ in the fermiophobic
limit of the 2HDM (type I)
was performed in \cite{Barroso:1999bf,Brucher:1999tx}.
The scalar potentials used in these references
contain one less parameter than that given in eq.~(\ref{scalar_pot}), and
thus their corresponding 
expression for $g_{h_fH^+H^-}$ differs from that given in 
eq.~(\ref{trilinear_coup2}).

In this work we perform a general scan of the 2HDM parameter space in order
to study the magnitude of $g_{h_fH^+H^-}$ and the impact of
the $H^\pm$ loops on  $B(h_f\to \gamma\gamma)$.
As input we use $\lambda_i$ ($i=1,2,3,4,5$) and $\tan\beta$. The mass 
parameters $m_{11}^2$ and $m_{22}^2$ in the potential are fixed by the 
minimization conditions in eq.~(\ref{tadpoles}). The third mass parameter
$m_{12}^2$ is related to the CP-odd Higgs mass $m_A$, as shown in 
eq.~(\ref{ma}), and fixed by the fermiophobic condition in 
eq.~(\ref{ferm_condition}).
The following perturbative limits for $\lambda_i$ are used:
\begin{eqnarray}
0 \,\, < & \lambda_1,\lambda_2 & < \,\, 4\pi/3 
\nonumber\\
-8\pi \,\, < & 2\lambda_3,2\lambda_4,\lambda_5 & < \,\, 8\pi
\label{pertlim}
\end{eqnarray}
In addition, the vacuum stability conditions for $\lambda_i$ 
given in \cite{Gunion:2002zf} are respected.

\begin{figure}[h]
\centerline{\protect\hbox{\epsfig{file=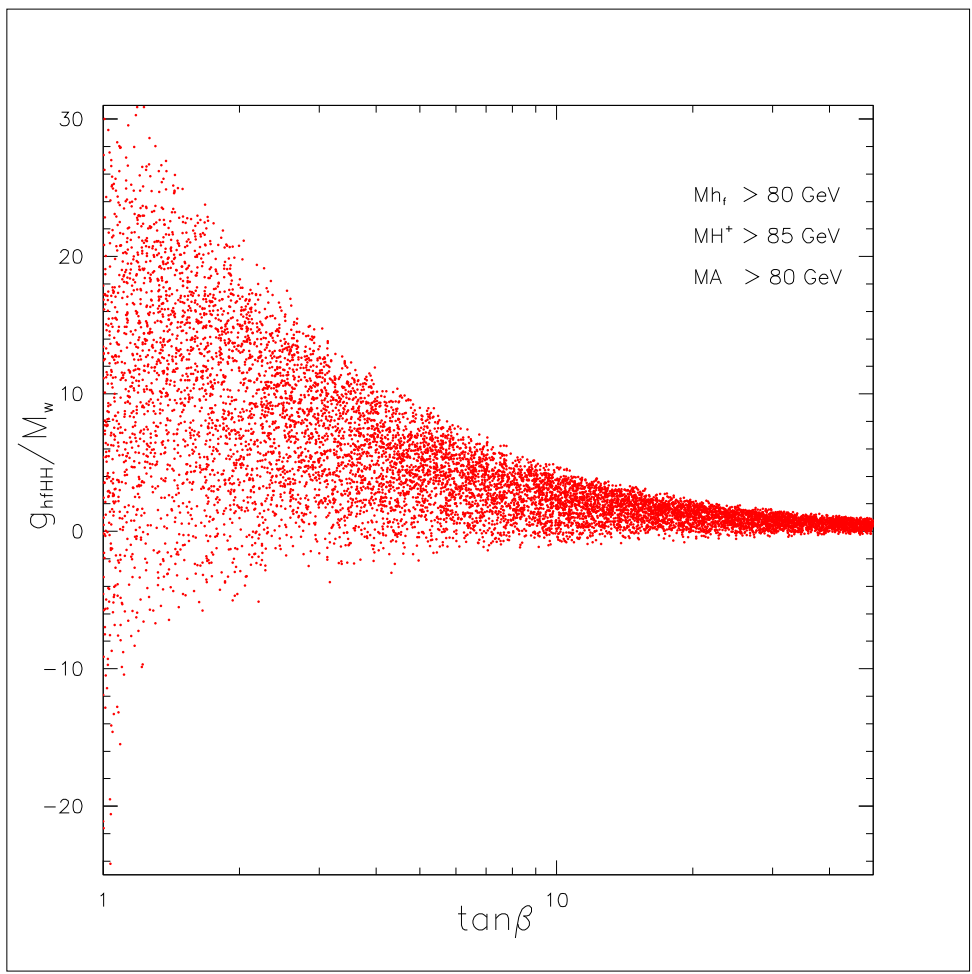,width=0.8\textwidth,angle=0}}}
\caption{\it Fermiophobic 2HDM $h_fH^+H^-$ coupling, normalized by $m_W$, 
as a function of $\tan\beta$.} 
\label{tbfree}
\end{figure}
In Fig.~\ref{tbfree} the magnitude of the trilinear $g_{h_fH^+H^-}$ coupling
[eq.~(\ref{trilinear_coup2}) normalized by $m_W$]
is shown as a function of $\tan\beta$.
We take $m_{h_f}, m_A \ge 80$ GeV, and $m_{H^+}\ge 85$ GeV in order to
comply with lower limits from direct searches. 
The $g_{h_fH^+H^-}$ coupling may have either sign, with the largest values
occurring for positive $g_{h_fH^+H^-}$.
Most strikingly, the absolute value of the coupling diminishes with 
increasing $\tan\beta$,
a fact that will have important consequences for the impact of
the $H^\pm$ loops on $B(h_f\to \gamma\gamma$).

\section{Numerical analysis}

In this section we perform a numerical analysis
of the effect of charged scalar loops on
B$(h_f \rightarrow \gamma \gamma)$. We also study the impact on 
the current searches for $h_f$ at the Tevatron Run II and on future searches
at the LHC. We consider both the standard production 
mechanism which depends on the $g_{h_fWW}$ coupling, and the complementary 
production mechanism which depends on the $g_{h_fH^{\pm}W}$ coupling.

\subsection{Searches for $h_f$ at the Tevatron Run II} 

Run II commenced in 2001 with the collision energy $\sqrt s$ increasing 
to 1.96 TeV. Simulations for the standard mechanism
$p\overline p\to V\to Vh_f$ ($V=W^\pm,Z$) 
can be found in Refs.\cite{Mrenna:2000qh,Landsberg:2000ht}. 
It was shown that lower limits of $m_{h_f}> 115 (125)$ GeV can be
obtained for the {\sl Benchmark F} model with 2 (10) fb$^{-1}$ of data, 
which is a moderate improvement over the LEP limits. 
However, the Tevatron search has the virtue of being sensitive to
the parameter space of $110$ GeV $< m_{h_f} < 160$ GeV, provided that 
$B(h_f\to \gamma\gamma)> 4\%$. In contrast, such 
a region could not be probed at LEP.
Although these large branching ratios are not possible
in the {\sl Benchmark F} model, we will discuss if contributions
from charged scalar loops ($H^\pm$) can provide 
the necessary enhancement.

A preliminary search for
$p\bar{p} \rightarrow W^*\rightarrow W h_f \rightarrow 
\gamma \gamma +X$
has been carried out with a Run II data sample
of 0.19 fb$^{-1}$ \cite{Melnitchouk:2005xv,D0result}.
Although the mass limit 
for $m_{h_f}$ in the {\sl Benchmark F} model 
is still inferior to that obtained 
at LEP, there is already sensitivity to 
the mass region $110 \,{\rm GeV} < m_{h_f} < 160
\,{\rm GeV}$ and $B(h_f\to \gamma\gamma)> 80\%$. As of June 2007,
2 fb$^{-1}$ of data have been accumulated.
In the fermiophobic limit in {\sl Benchmark A}
this production mechanism is suppressed  by a factor:
\begin{equation}
g^2_{h_fWW}\sim \cos^2 \beta = \frac{1}{1+\tan^2\beta}
\label{suppress_h_fWW}
\end{equation}
{\it i.e.}, at $\tan\beta=3$ there is already a suppression factor $1/10$,
and for this reason we will analyze this production mechanism at low values 
of $\tan\beta$. At larger values of $\tan\beta$ in {\sl Benchmark A}
one must rely on complementary production mechanisms.

Recently a search (with 0.83 fb$^{-1}$) for a complementary process
has been performed in the channel \cite{Landsberg:2007mc}:
\begin{equation}
  p\bar{p}\rightarrow h_f H^\pm\rightarrow h_f h_f W^\pm \rightarrow 
                              \gamma\gamma\gamma(\gamma) +X \; .
\label{comp_process}
\end{equation}
Such a mechanism has very little SM background and the absence of
a signal leads to the following limit on the production cross-section: 
\begin{equation}
\sigma(p\bar{p}\rightarrow h_f H^\pm) < 25.3 \,{\rm fb}
\label{yyy_limit}
\end{equation}
Eq.~(\ref{yyy_limit}) excludes regions in 
the parameter space of $m_{h_f}$-$\,m_{H^\pm}$ for a given $\tan\beta$
e.g. $m_{h_f}<44(50)$ GeV and $m_{H^\pm}<150$ GeV 
are excluded for $\tan\beta=3(30)$.
All the above searches assume {\sl Benchmark F} model $B(h_f\to \gamma\gamma)$.

Prospects for $ pp \rightarrow W^*\rightarrow W h_f 
\rightarrow \gamma \gamma +X$ at the LHC can be obtained
from simulations for the SM Higgs boson production mechanism 
$ pp \rightarrow W^*\rightarrow W \phi^0 
\rightarrow \gamma \gamma +X$ \cite{Lethuillier:2006bq}. 
For $m_{h_f}=120$ GeV in {\sl Benchmark F}, 
a simple scaling of the signal rates
in \cite{Lethuillier:2006bq} gives 
statistical signals of $20\sigma, 40\sigma$ and $70\sigma$ for 
${\cal L}=30,100,300$ fb$^{-1}$ respectively. For $m_{h_f}=150$ GeV,
$B(h_f\to \gamma\gamma)$ is approximately the same as
$B(\phi^0\to \gamma\gamma)$ and a $5\sigma$ signal 
can only be obtained with ${\cal L}=300$ fb$^{-1}$.
Simulations for the complementary channel in eq.~(\ref{comp_process})
have not yet been performed for the LHC. Given the low backgrounds,
one would expect considerably smaller values of
$\sigma(pp\rightarrow h_f H^\pm)$ to be probed than the
current upper limit set at the Tevatron Run II [eq.~(\ref{yyy_limit})].

\subsection{Impact of $H^\pm$ loops on $qq' \rightarrow W^*\rightarrow W h_f 
\rightarrow \gamma \gamma +X$}

\begin{figure}[h]
\centerline{\protect\hbox{\epsfig{file=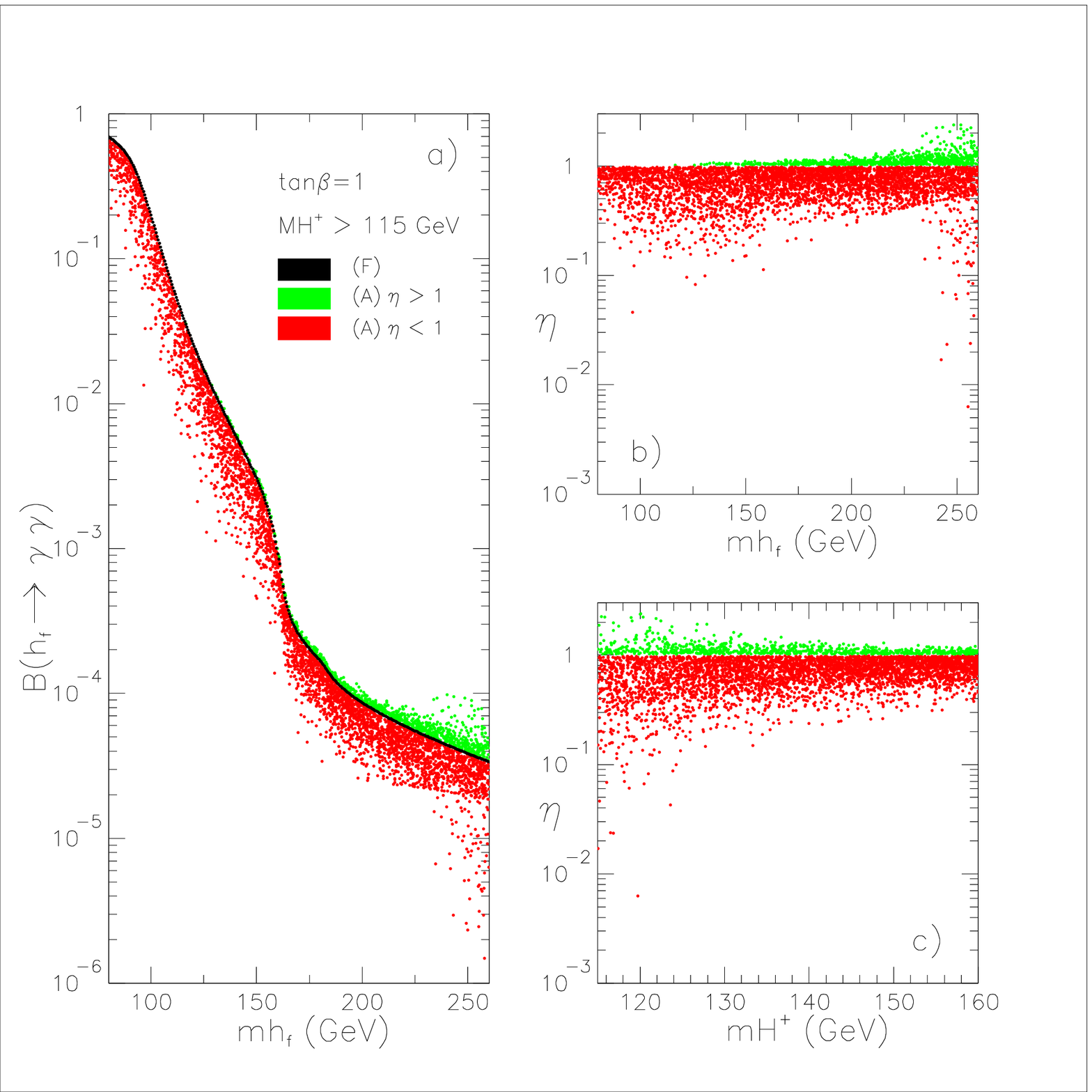,width=0.9\textwidth,angle=0}}}
\caption{\it For $\tan\beta=1$ the three panels are as follows: a) 
$B(h_f \rightarrow \gamma \gamma)$ in
 Benchmark F (black dots), and in Benchmark A with $\eta>1$ (green
 dots) and $\eta<1$ (red dots). 
b) Scatter plot for $\eta$ as a function of the fermiophobic Higgs mass
  $m_{h_f}$. c) Scatter plot for $\eta$ as a function of the charged Higgs 
mass $m_{H^\pm}$.} 
\label{tb1}
\end{figure}
In Fig.~\ref{tb1} we show the fermiophobic Higgs boson Branching Ratio 
$B(h_f\rightarrow\gamma\gamma)$ for $\tan\beta=1$ and $m_{H^\pm}\ge 115 $
GeV. The lower bound on $m_{H^\pm}$ ensures that the $H^\pm$
contribution to the decay $t\to H^\pm b$ \cite{Abulencia:2005jd} 
is consistent with the experimentally allowed regions.
In the case of $b\to s\gamma$, the current world average is 
$B(b\rightarrow s \gamma)=(3.55\pm 0.26)\times 10^{-4}$ 
\cite{Barberio:2006bi}. Strong lower bounds on $m_{H^\pm}$ 
independent of $\tan\beta$ can be derived from this decay 
in the context of the 2HDM (type II), but   
in the 2HDM (type I) of interest to us such bounds are
not applicable due to the different couplings of $H^\pm$ 
to fermions. For $\tan\beta>1$ the charged Higgs
contribution is small, and the prediction for $B(b\rightarrow s \gamma)$
approaches that of the SM as $\tan\beta$ increases. For $\tan\beta=1$ the
smaller values of $m_{H^\pm}$ used in this article predict  
$B(b\rightarrow s \gamma)$
somewhat lower than the experimentally allowed region, although this
theoretical prediction is unstable for $\tan\beta\le1$ due to scale 
dependence \cite{Borzumati:1998tg}.
In Fig.~\ref{tb1} {\sl Benchmark F} is denoted by the black points
while {\sl Benchmark A} is denoted by green and
red points for $\eta>1$ and $\eta<1$ respectively, with $\eta$ defined  
in the following way,
\begin{equation}
\eta = \frac{B(h_f \rightarrow \gamma \gamma)|_A}
{B(h_f \rightarrow \gamma \gamma)|_F}
\label{sup}
\end{equation}
which is the ratio of the branching ratios in {\sl Benchmarks A} and 
{\sl F}. In this way, for values of $\eta$ greater than unity, the 
charged Higgs and $W$ contributions add constructively. On the contrary, for
values of $\eta$ smaller than unity, the two contributions add destructively.

A fermiophobic Higgs boson decaying into two photons
was searched for at the Tevatron Run II with $0.19 \, {\mathrm{fb}}^{-1}$
\cite{D0result}.
The excluded region in the parameter space 
$\sigma(p\overline p\rightarrow h_f W) B(h_f\rightarrow\gamma\gamma)$ v/s
$m_{h_f}$ (Fig.~2 in \cite{D0result}) 
is marginally increased with respect to LEP. 
However, a Montecarlo prediction for 
$2 \, {\mathrm{fb}}^{-1}$ shows  a significant 
improvement over LEP. Assuming {\sl Benchmark F}, {\it i.e.}, no 
suppression factor in the cross section, and no charged Higgs contribution to 
the branching ratio, a fermiophobic Higgs mass $m_{h_f}\lsim 113$
GeV would be probed.
However, in {\sl Benchmark A} 
one has to include the suppression factor in the production 
cross section (eq.~\ref{suppress_h_fWW})
and the contribution of the charged Higgs boson to the Higgs
decay rate into two photons. In Fig.~\ref{tb1} we use 
$\tan\beta=1$ which gives rise to a suppression factor $g^2_{h_fWW}=1/2$. 
A few scenarios in 2HDM where the charged Higgs contribution to 
$B(h_f\rightarrow\gamma\gamma)$ overcomes the suppression factor 
({\sl i.e.}, $\eta > 2$) are observed in Fig.~\ref{tb1} for large values of
$m_{h_f}$ and small values of $m_{H^\pm}$. 
However, these large values $m_{h_f}\sim 250$ GeV together with 
the small $B(h_f\rightarrow\gamma\gamma)\sim 10^{-4}$
are well beyond the sensitivity of the Tevatron Run II.

In the mass region sensitive to Run II, $m_{h_f}\lsim 160$ GeV,
we find $\eta\lsim 1.1$ and thus the
constructive effects of the charged Higgs contribution to 
$B(h_f\rightarrow\gamma\gamma)$ are far from compensating the suppression
factor. Hence the use of {\sl Benchmark F} is reasonable for 
Run II when the search is negative and only lower bounds on the 
fermiophobic Higgs mass are set. We conclude that searches in 
the channel $qq' \rightarrow W^*\rightarrow W h_f 
\rightarrow \gamma \gamma +X$ with the anticipated Run II luminosity
of a few fb$^{-1}$ offer similar sensitivity to 
{\sl Benchmark A} as the LEP searches. 

At the LHC prospects are much brighter since 
statistically significant signals would be expected in 
{\sl Benchmark F} in the region $120 \,{\rm GeV} < m_{h_f} < 150\, 
{\rm GeV}$ (see Section 4.1).
If a signal were observed in the above mass region,
interpretation in the 2HDM (type I) (i.e. {\sl Benchmark A})
would require inclusion of the scalar loops, whose effect
on $B(h_f\rightarrow\gamma\gamma)$
can be sizeable as shown in Fig.~\ref{tb1}.
In the case of destructive interference for $m_{h_f}\sim 150$ GeV 
their contribution can be as large as $-90\%$ ($\eta\ge 0.1$);
in the constructive case, 
their contribution has an upper limit of the order of $10\%$ 
($\eta\le 1.1$) for the same value of $m_{h_f}$.
The signal event number in {\sl Benchmark A} would be
proportional to the cross section suppression
factor [eq.~(\ref{suppress_h_fWW})] and $\eta$. Information
on the magnitude of $\eta$ would restrict the parameter space
of the Higgs potential via eq.~(\ref{withscalarloops}) and  
eq.~(\ref{trilinear_coup2}). Since {\sl Benchmark F} can give very
large signals for lighter values of $m_{h_f}$
(e.g. 70$\sigma$ for $m_{h_f}=120$ GeV and 
300 fb$^{-1}$), even small values of $\eta$ could
be probed in {\sl Benchmark A} for $m_{h_f}\sim 120$ GeV. 

Note that the charged Higgs contribution
to $B(h_f\rightarrow\gamma\gamma)$ rises sharply near the threshold 
$h_f\rightarrow H^+H^-$. Since $m_{H^\pm}>115$ GeV in our example, the 
threshold starts appearing at 230 GeV. 
It is clear from Fig.~2c that the maximal value
for $\eta$ is obtained for light charged Higgs bosons.

\subsection{Impact of $H^\pm$ loops on
$qq' \rightarrow H^\pm h_f \rightarrow \gamma \gamma \gamma \gamma +X$ }

Complementary mechanisms play an important role in the search for $h_f$ in the
case of the $g_{h_fVV}$ coupling being suppressed, i.e., for large $\tan
\beta$. For this reason we will consider the 
process $ pp,p\bar{p} \rightarrow H^\pm h_f \rightarrow \gamma \gamma 
\gamma \gamma +X$ [eq.~(\ref{comp_process})].
The total cross section is given by
\begin{equation}
\sigma(pp,p\bar{p} \rightarrow H^\pm h_f)\times 
B(H^\pm \rightarrow W h_f)\times B^2(h_f \rightarrow \gamma \gamma)
\label{cscomp}
\end{equation}
The inclusion of the charged Higgs loops will affect the total
cross section quadratically, as can be seen in eq.(\ref{cscomp}), 
and thus their effect might be more important than for the standard mechanism.

\begin{figure}[h]
\centerline{\protect\hbox{\epsfig{file=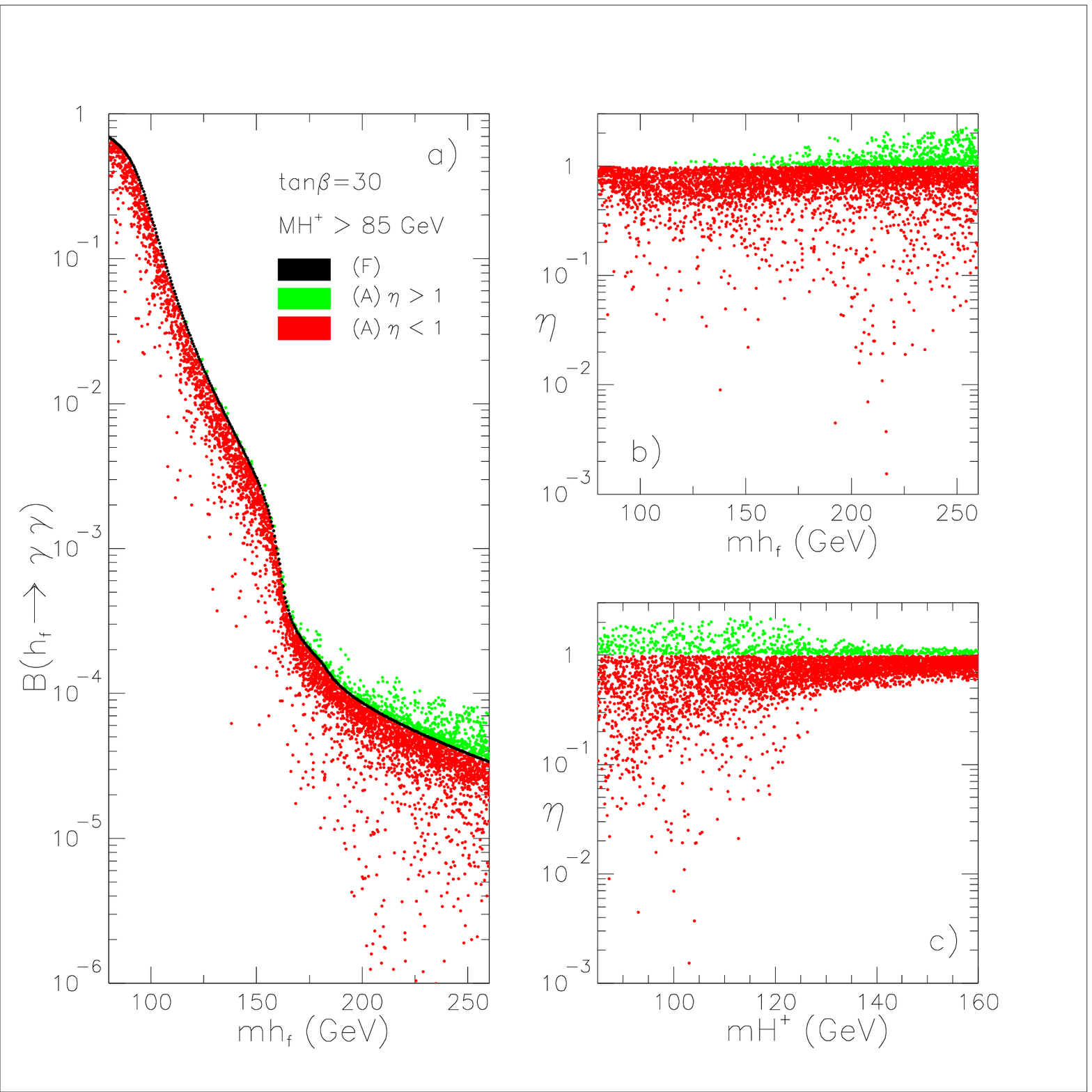,width=0.9\textwidth,angle=0}}}
\caption{\it For $\tan\beta=30$ the three panels are as follows: a) 
$B(h_f \rightarrow \gamma \gamma)$ in
 Benchmark F (black dots), and in Benchmark A with $\eta>1$ (green
 dots) and $\eta<1$ (red dots). 
b) Scatter plot for $\eta$ as a function of the fermiophobic Higgs mass
  $m_{h_f}$. c) Scatter plot for $\eta$ as a function of the charged Higgs 
mass $m_{H^\pm}$.} 
\label{tb30}
\end{figure}
In Fig.\ref{tb30} we show $B(h_f \rightarrow \gamma \gamma)$ as a
function of $m_{h_f}$ for $\tan \beta = 30$.
Since there is no cross section suppression factor
analogous to eq.~(\ref{suppress_h_fWW}), the signal event rate
in {\sl Benchmark A} will be enhanced relative to that
in {\sl Benchmark F} for $\eta>1$. In our scan of the 2HDM
parameter space we impose a weaker lower limit 
for the charged Higgs mass ($m_{H^\pm}>85$ GeV) than
in Section 4.2. This is because the rate for $t\to H^\pm b$ and the
$H^\pm$ contribution to $b\to s\gamma$ are
negligible for $\tan \beta = 30$ in the 2HDM (type I).
We plot both the constructive
(green dots) and destructive (red dots) contribution to 
the fermiophobic Higgs decay to two photons. 
In this figure, as in Fig.~\ref{tb1}, one sees a rise in the contribution to 
$B(h_f \rightarrow \gamma \gamma)$ (both constructive and destructive)
close to the charged Higgs threshold. In this case $m_{H^\pm}>85$ GeV 
and so the threshold starts to appear at 170 GeV.
In the mass region sensitive to the LHC, corrections from charged Higgs
bosons are very large; at $m_{h_f}\sim 150$ GeV they can
reach $25\%$ in the constructive case 
($\eta\le 1.25$), and $-99\%$ in the destructive case 
($\eta\ge 0.01$).

Since the $W$ contribution to $B(h_f \rightarrow \gamma \gamma)$ is
proportional to $\cos\beta$ [eq.~(\ref{withscalarloops})],
one might expect the influence of the charged Higgs contribution 
to be greater at large $\tan\beta$ when the $h_fWW$ coupling is very 
suppressed. However, from 
eq.~(\ref{trilinear_coup2}) one sees that in the fermiophobic limit, 
the $h_fH^+H^-$ coupling is also suppressed by a factor 
$\cos\beta$. For this reason the effect of the charged scalar loops is
comparable in both Fig.~\ref{tb30} and Fig.~\ref{tb1}, contrary to the
na\"ive expectation.

The lowest values of $\eta$ seen in Fig.~\ref{tb30} can be understood
as a cancellation between the $W$ and $H^\pm$ amplitudes contributing to 
the $h_f\rightarrow\gamma\gamma$ decay width, which in the notation of 
eq.~(\ref{withscalarloops}) translates into 
$F_0\,\tilde g_{h_fH^+H^-}=F_1\cos\beta$ for an exact cancellation. For 
the case with $m_{h_f}\approx 140$ GeV and $\eta\approx10^{-2}$, best seen 
in Fig.~4b, both amplitudes are real and have opposite signs, but the 
value of $\eta$ cannot be lowered significantly because the parameter 
$\lambda_1$ would exceed its perturbative limit shown in 
eq.~(\ref{pertlim}). For the case with $m_{h_f}\approx 220$ GeV and 
$\eta\approx10^{-3}$, both amplitudes have imaginary parts, since 
for this case $m_{H^+}\approx103$ GeV, and can be produced on-shell in
the decay of $h_f$. To obtain an exact cancellation between the $W$ 
and $H^\pm$ amplitudes in this case, it is necessary that the ratio 
of the real parts of $F_1$ and $F_0$ be equal to the ratio of their 
imaginary parts. It can be shown that this is not possible to achieve, and
that is the reason why we do not find smaller values of $\eta$.

\begin{figure}[h]
\centerline{\protect\hbox{\epsfig{file=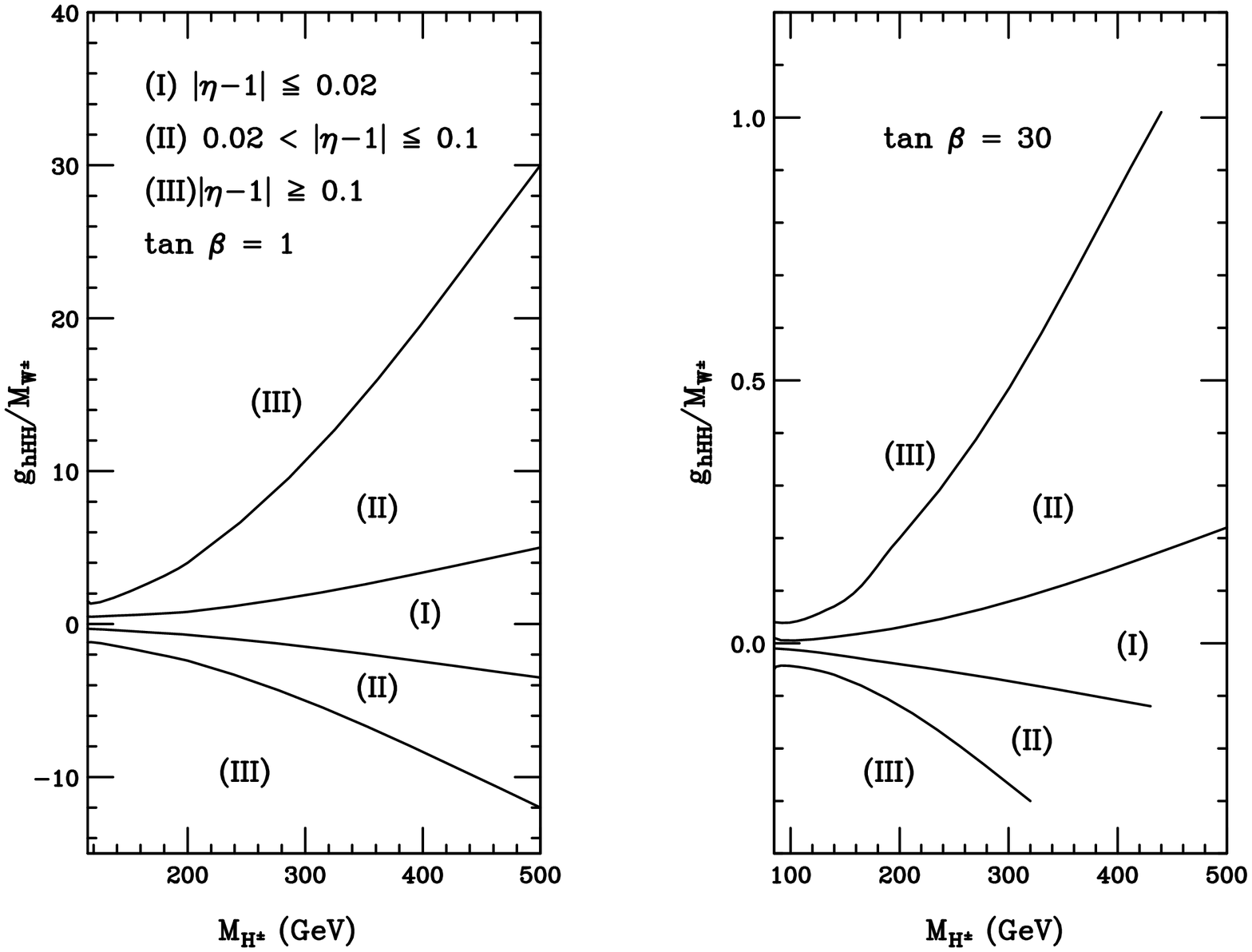,width=0.75\textwidth,angle=90}}}
\caption{\it In the plane of parameter space defined by the charged Higgs 
mass $m_{H^\pm}$ and the $h_fH^+H^-$ coupling $g_{h_fH^+H^-}$ (normalized by 
$m_W$), three different regions are displayed: I) Charged Higgs corrections 
within $2\%$, II) between $2\%$ and $10\%$, and III) larger than $10\%$, 
for two values of $\tan\beta=1, 30$.} 
\label{eta}
\end{figure}
In Fig.~\ref{eta} we show different regions 
in the plane formed by the 
$h_fH^+H^-$ coupling $g_{h_fH^+H^-}$, normalized by $m_W$, and the charged
Higgs mass $m_{H^\pm}$, for two values of $\tan\beta=1, 30$. The regions 
are defined by the parameter $\eta$ in the following way: I) 
for a given value of $m_{h_f}$,
$B(h_f \rightarrow \gamma \gamma)$ calculated in {\it Benchmark A} is 
within $2\%$ of the branching ratio calculated in 
{\it Benchmark F}; II) the deviation is between $2\%$ and $10\%$; III)
the deviation is larger than $10\%$. The dividing lines are not absolute,
since there is some small overlap between regions.
As expected, larger corrections are obtained for larger values of the 
coupling $g_{h_fH^+H^-}$. Similarly, larger corrections are obtained
for smaller charged Higgs masses. This behaviour can be understood from 
eq.~(\ref{trilinear_coup1}), where we see that for fixed $g_{h_fH^+H^-}$,
$\tilde g_{h_fH^+H^-}$ is inversely proportional to $m_{H^\pm}^2$.
The effect of $\tan\beta$ can also be
observed from the figure. The general shape of the regions is unchanged,
and the observable effect is a re-scaling of the coupling $g_{h_fH^+H^-}$.
This is clearly understood from eq.~(\ref{trilinear_coup2}), where we
see that both couplings $g_{h_fH^+H^-}$ and $g_{h_fW^+W^-}$ are scaled by 
a factor $\cos\beta$.

Note that the sign of the coupling $g_{h_fH^+H^-}$ directly determines
whether the charged Higgs contribution adds constructively ($\eta>1$)
or destructively ($\eta<1$) to the $W$ contribution. In
eq.~(\ref{trilinear_coup2}) this coupling is written as a function of 
$m_{H^\pm}$, $m_A$, $\lambda_1$, $\lambda_5$, and $\tan\beta$. The 
charged Higgs mass $m_{H^\pm}$ and $\lambda_1$ always contribute positively
to $g_{h_fH^+H^-}$. On the contrary, the CP-odd Higgs mass $m_A$
contributes negatively, while the $\lambda_5$ contribution does not have 
a definite sign. As stressed in Section 4.2, observation of
$h_f\to \gamma\gamma$ at the LHC with a sizeable event number 
would provide information on the magnitude of $\eta$, which in turn would
constrain the above Higgs potential parameters 
in the context of the 2HDM (type I).

At the LHC, with $100\,{\mathrm{fb}}^{-1}$ per experiment, the expected 
accuracy for the branching ratio $B(\phi_0 \rightarrow \gamma \gamma)$
varies between $14\%$ and $22\%$, within the mass region sensitive to
the LHC, $115<m_{\phi_0}<150$ GeV \cite{Zeppenfeld:2002ng}. Considering 
that this study was done based on the $gg\rightarrow\phi_ 0$ production
mechanism, it does not directly apply to our fermiophobic Higgs. This 
is because the relevant production mechanisms for $h_f$ at the LHC are 
higgsstrahlung 
and weak vector boson fusion, and a dedicated study would be needed. 
Nevertheless, it is encouraging that the corrections we find are easily 
larger than the quoted sensitivity. Even better precision for 
$B(\phi_0 \rightarrow \gamma \gamma)$ can be achieved 
at the ILC, being $16\%$ for $\sqrt{s}=500$ GeV, and improvable to $10\%$
with initial state polarization \cite{Boos:2000bz}, with higgsstrahlung
$e^+e^-\rightarrow Z\phi_0$ production mechanism, \ie, applicable to
a fermiophobic Higgs. Finally, the photon-photon option for the ILC
can achieve a $5\%$ and $8\%$ sensitivity for the modes 
$\gamma\gamma\rightarrow\phi_0\rightarrow W^+W^-$ and 
$\gamma\gamma\rightarrow\phi_0\rightarrow\gamma\gamma$, 
respectively \cite{Asner:2001vh},
which are also applicable to our fermiophobic Higgs $h_f$.

\section{Conclusions}

The LHC has impressive sensitivity to fermiophobic Higgs bosons ($h_f$) 
decaying to two photons with a large ($> 1\%$) branching ratio.
Observation of this photonic decay mode ($h_f \rightarrow \gamma \gamma$)
with a rate significantly above that expected for the 
Standard Model Higgs boson could
be accommodated in the fermiophobic limit of the 2HDM (type I).
In a commonly used benchmark model the decay 
$h_f \rightarrow \gamma \gamma$ is assumed to be mediated solely by 
loops involving $W^\pm$, although
potentially large contributions may arise from
charged scalar loops. In the mass region sensitive to the LHC
$(m_{h_f}<150$ GeV) we showed that such contributions can 
cause large suppressions, $\eta\ge 0.01$ ($-99\%$ correction) or moderate 
enhancements, $\eta\le 1.25$ ($25\%$ correction) of the branching ratio 
for $h_f \rightarrow \gamma \gamma$. These corrections should be compared 
with expected sensitivities in the measurement of the photon-photon 
branching ratio of the Higgs, which vary from $22\%$ at the LHC to
$5\%$ at the photon-photon option of the ILC.
Consequently, interpretation in the 2HDM (type I)
of any signal for $h_f$ at the LHC and ILC would necessitate inclusion 
of the scalar loops. This in turn would provide information on the 
parameters of the scalar potential, through the charged Higgs mass and the 
$h_fH^+H^-$ coupling, which in the fermiophobic limit was shown 
to diminish with increasing $\tan\beta$, as does $h_fW^+W^-$.

\section*{Acknowledgments}  
A.G.A. was supported by National Cheng Kung University Grant 
No. OUA 95-3-2-057.
M.A.D. was supported by Anillo ``Centro de Estudios Subatomicos'' grant.
M.A.R. is thankful to Conicyt for their support.
A.G.A. is grateful for travel support from Fondecyt project No. 7060202 
and for hospitality at La Universidad Cat\'olica de Chile where this
work was initiated.

\end{document}